\documentclass[manuscript]{aastex}

\usepackage{epstopdf}
\usepackage{subfig}
\slugcomment{Based on observations collected at the Centro Astron\'omico Hispano
Alem\'an (CAHA) at Calar Alto, operated jointly by the Max-Planck
Institut f\"ur Astronomie and the Instituto de Astrof\'{\i}sica de
Andaluc\'{\i}a (CSIC),
at Lulin observatory operated by the Institute of Astronomy, National Central University in Taiwan, and at Xinglong station inaugurated by National Astronomical Observatory (BAO), Beijing }

\shorttitle{103P activity}
\shortauthors{Lin et al.}

\def\prod{$\rm s^{-1}$}

\begin{document}


\title{Long-Term Monitoring of Comet 103P/Hartley 2}

\author{Z.-Y. Lin}
\affil{Institute of Astronomy, National Central University
No. 300, Jhongda Rd, Jhongli City, Taoyuan County, Taiwan, 32001}
\email{zylin@astro.ncu.edu.tw}

\author{L.M. Lara} 
\affil{Instituto de Astrof\'{\i}sica de Andaluc\'{\i}a (CSIC),
       Glorieta de la Astronom\'{\i}a s/n, ES-18008 Granada, Spain}
\email{lara@iaa.csic.es}

\and

\author{W.-H. Ip}
\affil{Institute of Astronomy, National Central University
No. 300, Jhongda Rd, Jhongli City, Taoyuan County, Taiwan, 32001}


\begin{abstract}

We reported the monitoring results on spectrophotometry, photometry and imaging of comet 103P/Hartley 2 obtained at Lulin (1m), Calar Alto (2.2m) and Beijing Astronomical (2.16m) Observatory from April to December 2010. We found that a dust feature at sunward direction was detected starting from the end of September until the beginning of December (our last observation from the Lulin and Calar Alto observatory). Two distinct sunward jet features in the processed images were observed on October 11 and after October 29 until November 2. In parallel, the CN images reveal two asymmetrical jet features which are nearly perpendicular to the Sun-nucleus direction and this asymmetrical features implies that the comet was in a nearly side-on view in late-October and early-November. Additional to the jet features, the average result of the C$_2$-to-CN production rate ratio ranges from 0.7 to 1.5 which places 103P/Hartley 2 as being of typical cometary chemistry. We found that the $r_h$ dependence for the dust production rate,  Af$\rho$  (5,000 km), is $-$3.75$\pm$0.45 before perihelion and is $-$3.44$\pm$1.20 during post-perihelion period. We detected the higher dust reddening is around the optocenter and getting bluer outward along the sunward jet feature and concluded that the former one, higher dust reddening, could be associated with strong jet activity and the latter one, the lowering of the reddening, might imply the optical properties changed or could be associated with outburst.  The average dust color did not appear to vary significantly as the comet passed through perihelion.
\end{abstract}

%%Steeper slopes may be caused by non-steady state emission, or by fading grains, whereas shallower profiles require a source function in the coma (e.g., fragmenting grains or production of gas from grains). 
%%\textbf{to conclude this, you have to do the analysis on images at different rotational phases, something which is not show at all in the manuscript. We should "label" the images with rotational phases too and I believe that we will see color variations, at least in the inner coma}




%%  Otherwise, LaTeX will issue an error

\section{Introduction}

Comet 103P/Hartley 2, hereafter referred to as Hartley 2, was first spotted by M. Hartley on March 16, 1986. It has a semi-major axis of $\rm a =3.47 AU$, eccentricity $\rm e=0.695$, and inclination $\rm i = 13.617$ and an orbital period of 6.46 years. Its low eccentricity made it a suitable target for the extended mission of NASA's Deep Impact spacecraft after the impact experiment at comet 9P/Tempel 1 on July 4, 2005. The mission to Hartley 2 was renamed EPOXI and given two missions, Extrasolar Planet Observation and Characterization(EPOCh), and Deep Impact Extended Investigation (DIXI). The EPOXI flyby observations at a closest distance of 694 km on November 4, 2010, brought a wealth of information on the outgassing activity, shape and surface structure of this small Jupiter family comet (A'Hearn et al., 2011). For example, the strong outflows of the CO$_2$-rich jet from the sun-lit end of the bowling-pin shaped and the H$_2$O-rich jet in the waist region came as a surprise. How would they be connected to the large-scale jet structures observed in the coma? How would the outgassing process be modulated by the rotation of the comet nucleus? In fact, based on the time variability of the CN coma morphology and millimeter/sub-millimeter spectra, the rotation period of Hartley 2 has been found to be increasing from 16.7 hr in August, 2010, to 18.4 hr in the first half of November and then to nearly 19 hr in late November \citep{sam11, kni11, mee11,wan12}. Such time variations of the nucleus rotation period together with the close-up measurements by the EPOXI mission demonstrate the complex nature of the surface outgassing process. 

In anticipation of the scientific opportunity to compare the large-scale coma structures and gas production rates of Hartley 2 with the EPOXI results, we have made a long-term monitoring program from April to December, 2010, using imaging with both broadband and narrowband filters, and long-slit spectrophotometry. This cooperative effort involved observations at the Lulin Observatory in Taiwan, the Calar Alto Observatory in Spain, and the Beijing Astronomical Observatory in China. The paper is organized as follows. In Section 2, we will explain the observational procedures, instruments and analysis methods. In Section 3, the derived morphology and gas production of the CN coma and jets will be described. In Section 4, we will describe the dust jets and the structure of the dust coma during this period. A summary of the major characteristics of the large-scaled structures of the gas and dust comas of Harley 2 is given in Section 5. 

\section{Observations, instruments and data analysis}

{\bf Imaging:} The bulk of the photometric imaging observations was done by using the Lulin One-meter Telescope  (LOT) at Lulin observatory. In our first image of Hartley 2 on April 24, 2010, when the comet was 2.42 AU away from the Sun and 2.36 AU from the Earth, only a diffuse coma of 5" diameter was visible with 10-min exposure time. There was no tail feature. In the monitoring program, an Asahi R broadband filter and the narrowband filters of the Rosetta filter set were used. The specifications of these narrowband filters are given as $\lambda_{c}/\Delta \lambda$ both in nm, $\lambda_c$ being the central wavelength and $\Delta \lambda$ the band width:  CN (387/5nm), C$_{2}$ (512.5/12.5nm), blue continuum BC (443/4nm), and red continuum RC (684/9nm). Because of the consideration of the signal-to-noise ratio, the narrowband filters were used only in October and November 2 just before the EPOXI close encounter. The camera used on LOT from April to November was PI 1300B which has a pixel scale of 0.516 arcsec and a field of view of $11.2 \times 11.6$ arcmin. In late-November 2010, there was a cooling problem with PI1300B. We, therefore, switched to U42 CCD which has $\rm 2k \times 2k$ pixels and a field of view of  $12.17 \times 11.88$ arcmin. The telescope was always operated with non-sidereal tracking so as not to produce trail in the comet images. Typical integration were 600s $\sim$ 900s for the narrowband filters and 30s  $\sim$ 300s for the broadband R filter.

Table 1 is the observational log of our program. Standard procedure of data reduction was applied. It began with dark current subtraction and flat-field correction of all image frames. This was then followed by the subtraction of the night sky contribution. For the observations obtained before late-September, the night sky levels were determined directly areas of the CCD frames that do not contain contributions from the cometary emission. However, the sky-background of those images taken from late-September to early-November was all influenced by the cometary coma. Therefore, we took sky background images positioned at about 0.5 degrees away from the comet center. The extinction coefficients of the narrowband and broadband R filters were determined for all nights with photometric sky conditions, using the photometric stars like Feige 110 and GD71, observed at different airmasses during the night. For example, the first order extinction coefficient (in units of magnitudes per air mass) measured by Feige 110 with observing airmass range from 1.1 to 1.7 for R-filter on October 29 is 0.10 and for CN and C$_2$ are 0.39 and 0.14, respectively. These data were used to convert the measured counting rates into physical units and the detail has been described in Lin et al. (2007b). 
Because the CN images contain $29\%$ contribution from the continuum in the blue range while the C$_2$ images have as much as $93\%$, the net CN and C$_2$ gas coma images need to go through the subtraction procedure according to the following formulas: CN= CN$_{obs}$ - 0.29 BC$_{obs}$ , C$_2$ =  C$_2^{obs}$  - 0.93 BC$_{obs}$. 

In addition to Lulin observations, the coma activity of 103P/Hartley 2 was also monitored continuously in R-band from the Calar Alto Observatory (near Almeria, Spain) from July 14 to December 26, 2010 (see Table 1). We used the CAFOS imaging camera ($\rm 2k \times 2k$ pixels, pixel size: 0''53,  FOV $18' \times 18'$) which was mounted on the 2.2 m telescope. In our observations, only the central 1k x 1k pixels were used, providing thus a FOV of $9' \times 9'$.  Appropriate bias and flat field frames were taken each night. If photometric conditions prevailed, photometric standard stars were observed at airmass similar to the comet observations. Table \ref{ObsLog} contains the observations log for the complete dataset. Notice that a larger number of night mentioned in Table \ref{ObsLog} is for the Af$\rho$ estimation and some of them are not used to enhance the structures in the coma because the SNR was too low. 

 {\bf Spectroscopy:} Spectroscopic measurements were planned once every month using CAFOS with grism B400 (see http://www.caha.es/alises/cafos/cafos22.pdf) which renders an spectral range between 3,200 and 8,800 \AA\ with a wavelength scale of 9.4 \AA/pixel. The slit of the spectrograph was oriented in the north-south direction, giving dust and gas profiles at different cross-cuts through the coma, depending on the PA of the sun-comet vector on the sky. For absolute calibration, observations of appropriate spectrophotometric standard stars were acquired. All comet observations were done with telescope tracking on Hartley 2. With the exception of Nov. 5, 2010, all observations were done in service mode of the Calar Alto Observatory. Details on the images and spectra reduction and calibration can be found in Lara et al. (2001, 2011a)  and they will not be repeated here. If the gas coma covered the whole slit, the sky level was estimated from the edges of the frame. Otherwise, the background could be measured directly by using regions near the edges of the frame.

Besides spectra obtained from the Calar Alto Observatory, spectroscopic observations were also performed on October 9 and October 11 at the Beijing Astronomical Observatory using the 2.16m telescope in the spectral range between 3,600 \AA $\smallskip$  to 8,400 \AA $\smallskip$  at a dispersion 4.8 \AA/pixel. The spectroscopic data were reduced following the standard procedures including bias and flat-field corrections and cosmic ray removal. Wavelength calibration was performed based on helium-argon lamps exposed at both the beginning and the end of the observations every night. Flux calibration of all spectra were conducted based on observations of at least one of the spectral standard stars, i.e., HD19445 and the atmospheric extinction effect was corrected by the mean extinction coefficients measured by the Beijing-Arizona-Taiwan-Connecticut (BATC) multicolor survey. See Lin et al. (2007a) for more detailed information. 
 
\section{Gas coma morphology and properties}
\subsection{CN jets}

In order to study the visibility of faint structures of the gas coma of comet Hartley 2, an image enhancement technique was applied to the present set of images. The method used here is the azimuthally averaged profile division, a detailed description which can be found in Lin et al. (2012). This method was applied to all images taken in the CN filter, in continuum filters and in the R-band filter. 

To estimate the rotational phase from CN morphology, a lot of observing data have to be acquired in consecutive night. However, the images obtained in our observing nights with less temporal coverage were not enough to estimate and display rotational period due to snapshot observations, poor weather and telescope tracking problem. We, therefore, use the known periodicity to estimate the rotational phase in our images. However, we have to face several problems: a non-principal axis rotation of comet Hartley 2 and a rapid change of the viewing geometry might cause different periodicities between rotational cycles. A specific phase is really only applicable to a short stretch of data if we adopt known periodicity such as 18.15 hr in mid-October and 18.7hr in early-November from Knight and Schleicher (2011) or 18.22 hr around perihelion from Harmon et al. (2011). Notice that those ground-based observations have error bars between 0.01 and 0.3. The most robust rotation period at present is from the EPOXI spacecraft lightcurve given in Belton et al. (2013). This gives a spin period of 18.40 hr at encounter and states that it was increasing by 1.3 minutes/day. As the data acquired with the Rosetta filter set spread around one month, it is appropriate to use the midpoint of the observational time interval for this period of time. We extrapolate the rotation period back to midpoint assuming the rotation period was steadily changing during this time frame. Therefore, the rotation period quoted in this work is 18.11 hr on October 21.5 UT which refers to the midpoint of the Oct 10-Nov 2 data. The zero phase is set at 11:40 UT on October 10, 2010.

%%The phases changing from 0 to 1 are obtained using the rotational period of 18.2 hr from October 10 to 27 and 18.7 hr from October 28 to November 2, respectively (Knight and Schleicher 2011 and Harmon et al. 2011). The zero phase is set at 11:40 UT on Octorber 10, 2010.

\begin{figure}[p!]
\begin{center}
\includegraphics[width=\textwidth]{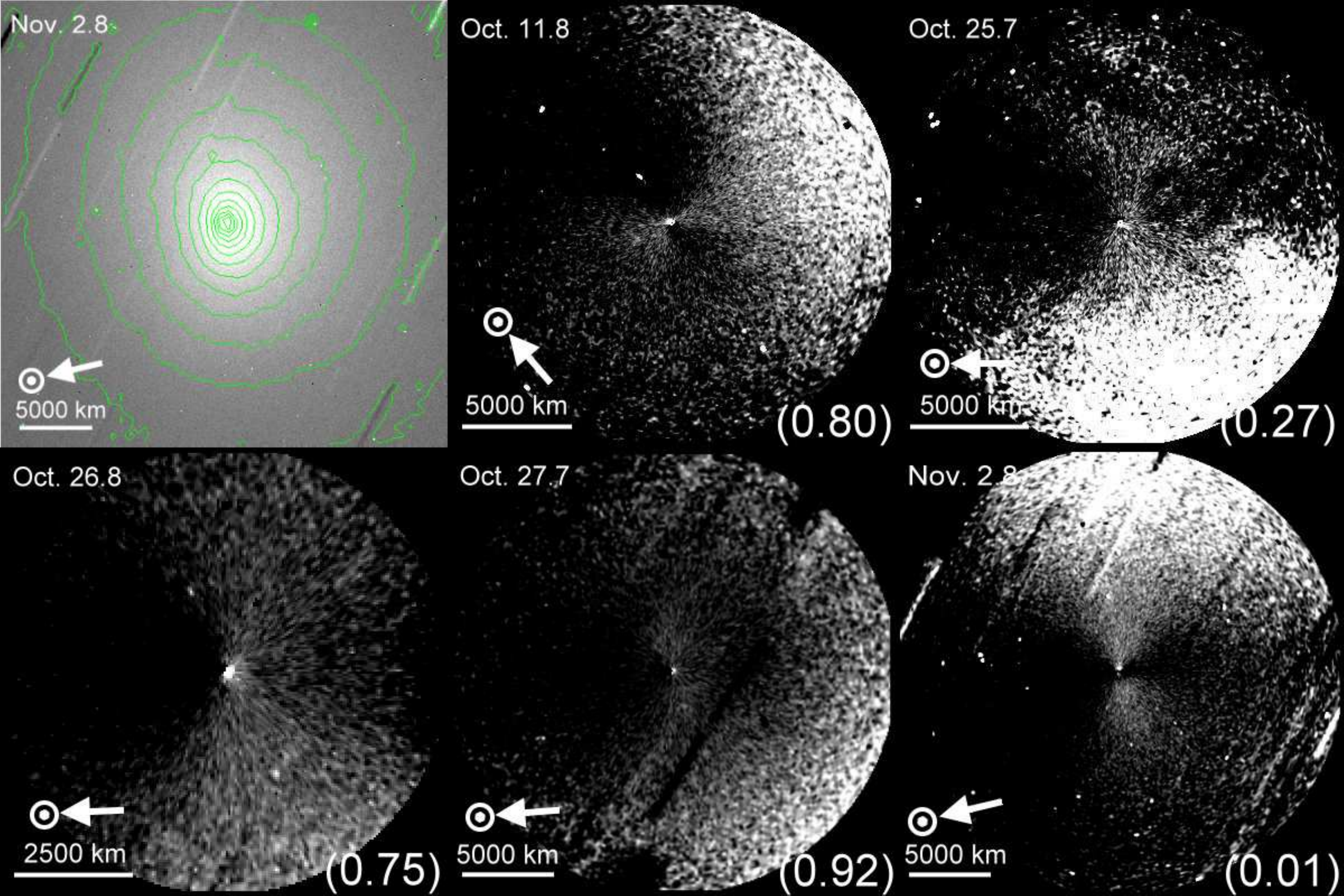}
\caption[]{
	\label{CNjet}
	 The CN images after the dust continuum was removed were enhanced by dividing by an azimuthal median profile. The rotational phase is given in the bottom-right corner of each image (see text for details). The original image (left-top) obtained on November 2 is shown with contours overplotted in green. The Sun symbol and arrow indicate the projected direction towards the Sun. North is up, East is to the left. The field of view is 3.44' $\times$ 3.44' and the scale bar is shown at the bottom left corner. The images are centered on the optocenter and the color code stretches for white representing the brightest areas and for black representing the darkest areas. 
	}
\end{center}
\end{figure}

In Figure ~\ref{CNjet}, we can see that the morphology of the CN coma extended almost perfectly along the east-west direction in early October and the north-south direction around Hartley 2's perihelion. The CN images all showed clear asymmetries before performing the image enhancement. One of these unprocessed CN images is given by contour plot in Figure 1 (top-left panel). The variations in between early-October and around its perihelion in the CN jet features are related to the spin period of the comet nucleus, the changing viewing geometry and non-principal axis rotation as has already been reported by Samarasinha et al. (2011), Knight and Schleicher (2011), Lara et al. (2011b) and Waniak et al. (2012). The processed CN images from the observations between October 11 and November 2 revealed two jets in the coma of comet Hartley 2. The CN jet features being nearly perpendicular to the Sun-tail direction not only varied smoothly during a night but showed similar morphology near its perihelion even though the rotational state was different. 

We compared the morphology of the CN jet features with those presented by Knight and Schleicher (2011), and Samarasinha et al. (2011) and found that the CN jet features of Hartley 2 did not show the spiral-like structure in early-October but was compatible with the observations obtained by Knight and Schleicher (2011) and Lara et al. (2011b)  in late-October. The reason could be the observing geometry, i.e., whether it is observed from the face-on or side-on. Knight and Schleicher (2011) confirmed this effect from the images that revealed the face-on spiral structures in August and September. Furthermore, we found these two CN jet features to be asymmetrical. One of them is always brighter than the other, possibly because it is facing toward the Earth. For example, the southern jet of the images obtained from October 25 to October 27 is slightly stronger than the northern jet of those images. Such asymmetrical features have also been reported in earliler works by  Samarasinha et al. (2011), Lara et al. (2011b) and Waniak et al. (2012).

\subsection{Gas production rates}

In order to determine the gas production rates, the mean radial emission profiles of CN and C$_{2}$ were derived from the images with the continuum subtracted. Regarding the spectra of the comet acquired at Calar Alto Observatory and Beijing Astronomical Observatory, they are also used to investigate the CN, C$_{3}$, C$_{2}$ and NH$_{2}$ profiles in the North-South direction and to derive the production rates of these gaseous species. The spectral regions and the subtraction of the underlying continuum in the gas emission bands were done as described by Lara et al. (2001). The conversion of the emission band fluxes into column densities made use of fluorescence efficiency factors ($g-$factors) for C$_{3}$, C$_{2}$ and NH$_{2}$ \citep{ahe95}, whereas the  $g-$factors of the CN molecule was calculated for the heliocentric distance and velocity of 103P/Hartley 2 on every date from the set of values given by Schleicher \citep{sch10}. The gas production rates are obtained by means of the Haser \citep{has57} model for isotropic emission of cometary neutral molecules and their daughter molecules and radicals. The parameter used for the parent velocity is  v$_{p}$ = 0.85 $r_{h}^{-0.5}$ $ \rm km  s^{-1}$ \citep{fra05} and for the daughter velocity it is 1 $\rm km  s^{-1}$. For the corresponding set of parameters in the Haser model, we produced theoretical column density profiles for each species by varying the production rate until the best match between observations and theoretical predictions is achieved. The results of nightly averages for Q(CN), Q(C$_3$), Q(C$_{2}$) and Q(NH$_2$) are summarized in Table~\ref{Qs}. Table \ref{Qs} also contains the average gas production rates obtained from the images acquired in one night together with the aperture size we have considered to derive Q. The variation of production rates seen in multiple measurements during a night were less than 5\% that is reflected in the uncertainties in Table ~\ref{Qs}. Our results on Q(C$_{2}$) are less numerous as there were tracking problems at LOT from October 10$\sim$11 and October 25$\sim$27, whereas the long-slit spectroscopic measurements could provide Q(C$_2$) at other dates thus spanning larger heliocentric distances.  

\begin{figure}[p!]
\begin{center}
\includegraphics[width=\textwidth]{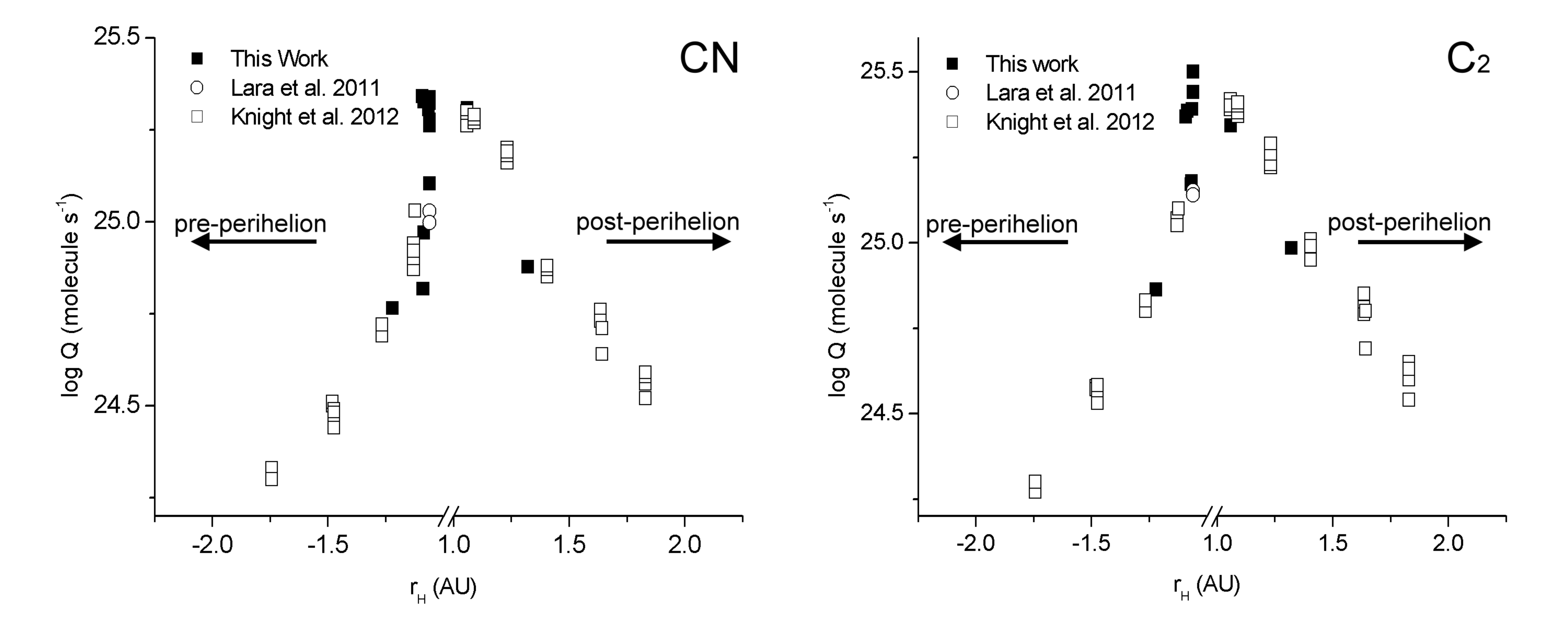}
\caption[]{
	\label{CNC2}
	 Log of production rates for CN (left) and C$_{2}$(right) plotted as a function of the heliocentric distance. Different symbols come from different data sets: filled square symbols refer to results here presented; open square symbols are taken from results in Knight and Schleicher (2013) and the open circle symbols come from Lara et al. (2011b). "//"is  referred to the break heliocentric distance from -0.1 AU(pre-perihelion) to 0.99 AU(post-perihelion). }
\end{center}
\end{figure}

Our Lulin, BAO and CA results  in Table~\ref{Qs} show that there is no significant variation of Q(CN) from mid-October to early-November. This result is consistent with the Lara et al. (2011b) and Mumma et al. (2011) results that assumed that HCN is the main parent species of CN and that expected variation of Q(HCN) around the perihelion is not very large. Notice that we used the mean radial profile to estimate the gas production rate from the images obtained from Lulin observatory. However, if we averaged the radial profile in the north-south direction where the CN jet feature exists, the derived Q(CN) would be larger in a factor of two to three when compared with the azimuthally averaged radial profile. Figure  ~\ref{CNC2} shows the logarithm of the production rate for CN and C$_{2}$ as a function of the heliocentric distance (r$_{h}$). The data points include those obtained by Lara et al. (2011b), Knight and Schleicher (2013) and the Lulin and CA results (this work) for pre-perihelion and post-perihelion observations during the 2010 apparition are presented here. We used the linear fitting in the log-log scale to estimate the slope of the r$_{h}$ dependence of the gas production rate, Q$\rm \sim r_{h}^{ -\alpha}$, and the slopes ($\alpha$)  of CN and C$_{2}$ are 4.57 and 4.84 before perihelion and 3.21 and 3.42 after perihelion, respectively. The corresponding  slopes are significantly steeper than the average value estimated for Jupiter-family comets, i.e. Q(gas)$\rm \sim r _h^{-2.7}$ \citep{ahe95}. Additionally to this, the average C$_2$-to-CN production rate ratio is $0.7 \sim 1.5$ which places 103P/Hartley 2 as being "typical" in terms of cometary chemistry defined by A'Hearn et al (1995) . Our measurement is consistent with the results from the spectroscopic observations (Lara et al. 2011b) and the narrow-band photometry observations (Knight and Schleicher 2013).

\section{Dust coma morphology and properties}
\subsection{Jet feature in dust coma}

\begin{figure}[p!]
\begin{center}
\includegraphics[width=\textwidth]{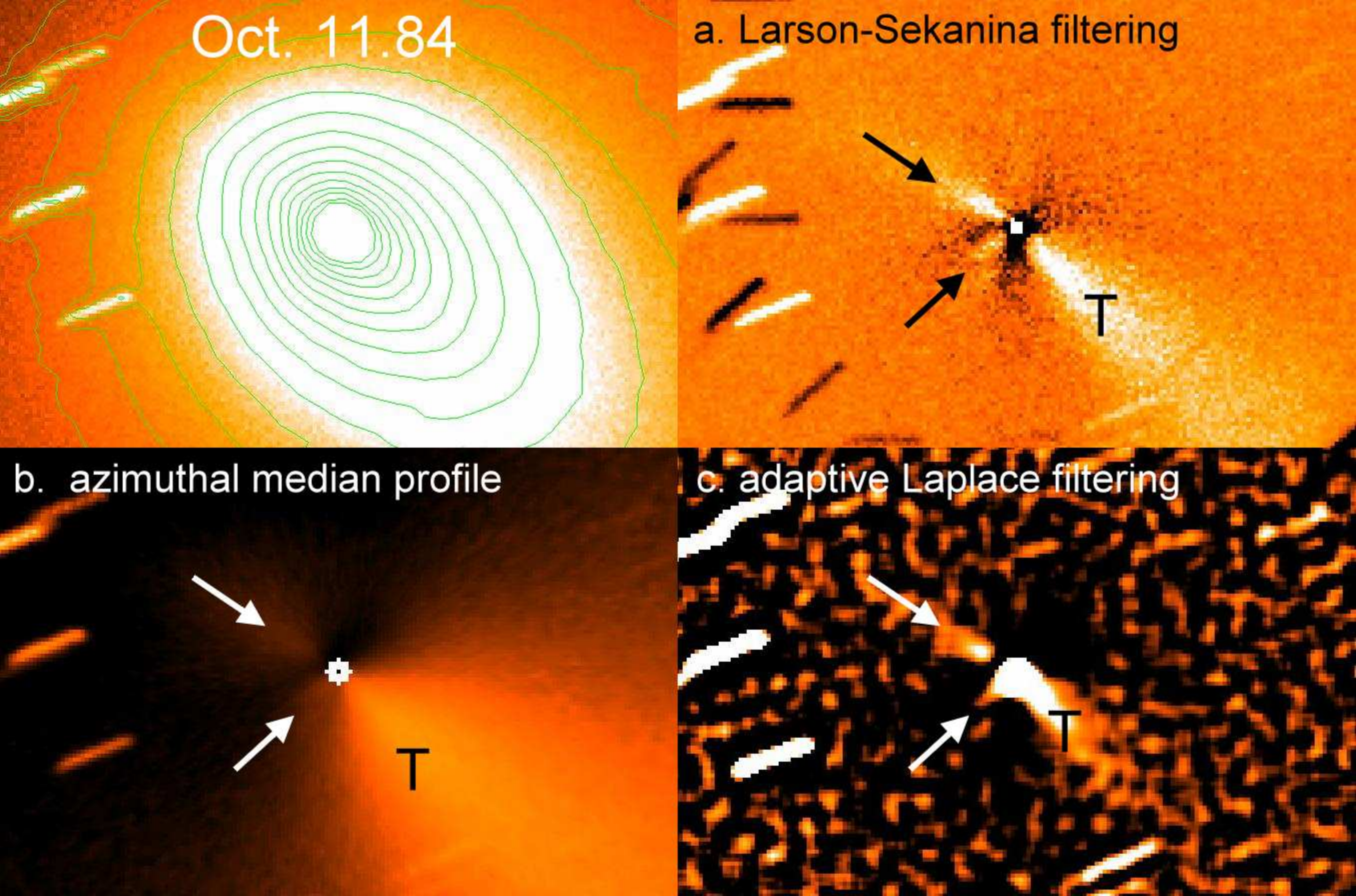}
\caption[]{
	\label{comparison}
	Image of comet 103P/Hartley 2 obtained on October 11, 2010 with R broadband filter. At the top left corner, a contour plot of the original image is shown. In (a) we display the same image after Larson-Sekanina filtering, in (b) the image is divided by an azimuthal median profile, and in (c) the adaptive-Laplace technique has been applied. In all of them, two jet features are visible. North is up, East is to the left., the field of view is 2.92' $\times$ 1.94', corresponding to  9,200 x 6,100 km at the comet distance. The images are centered on the nucleus, the arrows point out the jets, T labels the tail, and the streaks are trailed stars. The negatives of the star trails in panel A are the artifacts of the resulting image subtracted using a combination of a 15$^\circ$ counter-clockwise rotation and a 15$^\circ$ clockwise rotation. As the images are normalized, the brightness scales from 0.95 to 1.05.
	}
\end{center}
\end{figure}

We describe the morphology and evolution of the coma structures that can be treated with routine procedures, i.e. Larson-Sekanina algorithm (Larson and Sekanina 1984).  In case of doubt, we used additional techniques, such as azimuthal median profile division and Adaptive Laplace filter (B\"ohnhardt and Birkle 1994) to clearly separate morphological features from artifacts. Figure~\ref{comparison} compares the jet structure and dust tail feature on October 11 obtained by using three different image enhancement methods: (a) the Larson-Sekanina filtering,  (b) the azimuthal median profile, and (c) the adaptive Laplace filtering.  In spite of some differences in their appearances, the presence of two jets in the sunward quarter is common to all numerical treatments. It is therefore clear that the jet features are real and not artifacts associated with the image processing procedures. 

\begin{figure}[p!]
\begin{center}
\includegraphics[width=0.7\textwidth]{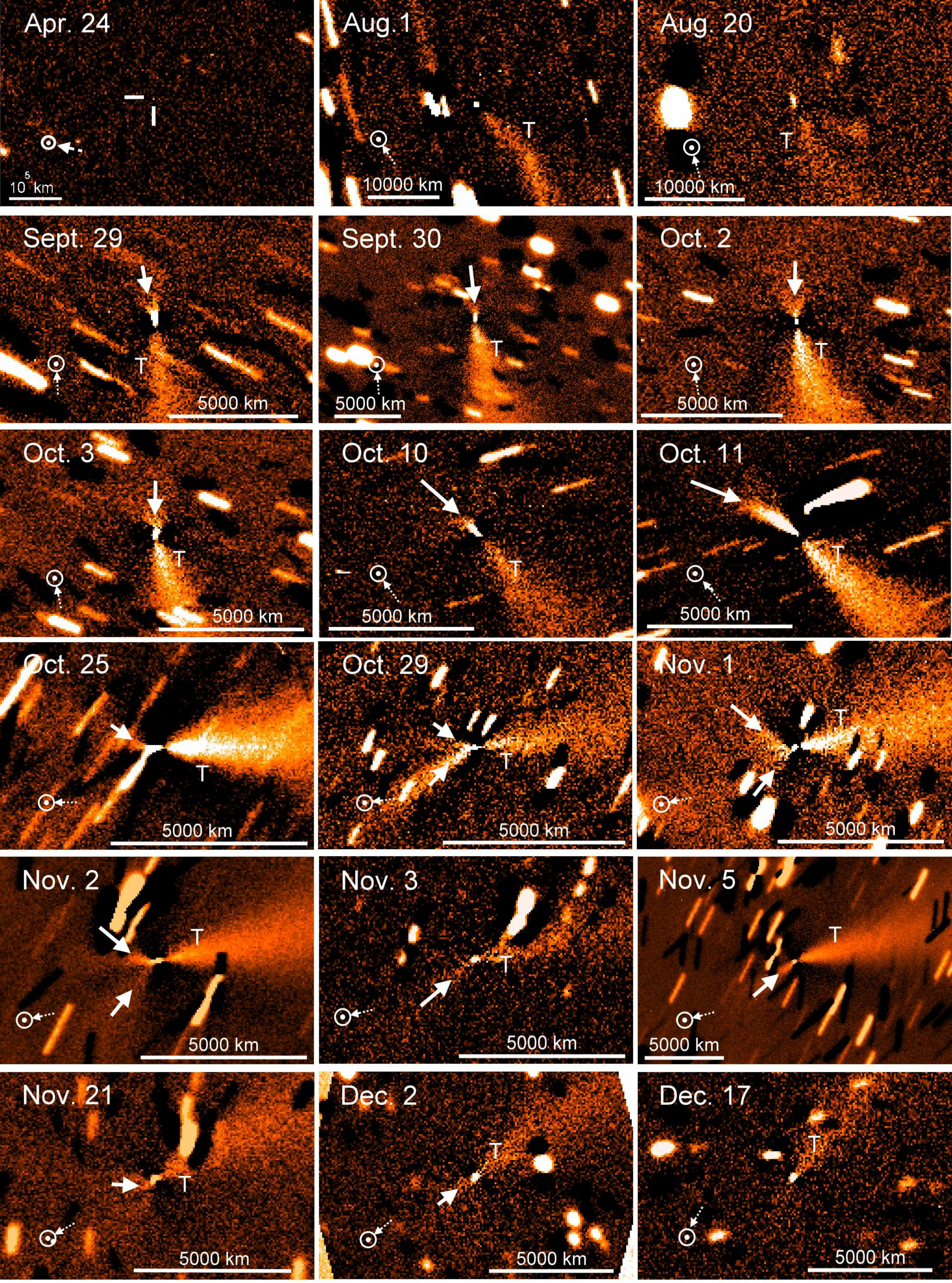}
\caption[]{
	\label{Rjet1}
	Images of comet 103P/Hartley 2 acquired with the R-band images and enhanced by Larson-Sekanina algorithm. The dust sunward jet feature sometimes represents a straight jet but sometimes it shows the multiple jet features during a night.  The jet showed minimal change in shape, position angle and extent from night to night. The Sun symbol and arrow indicate the projected direction towards the Sun. North is up, East is to the left. The field of view is 2.92' $\times$ 1.94' and the scale bar is shown in the bottom corner. All images are centered on the nucleus, arrows point out the jets, and T represents the tail.
	}

\end{center}
\end{figure}

Figure~\ref{Rjet1} is a summary of the R-band images enhanced by the Larson-Sekanina filtering method to bring out the inhomogeneous structures in the dust coma of 103P/Hartley 2.  It can be seen that from April until July, 2010, no clear sign of dust features could be found. However, beginning in August 1, a dust tail of diffuse structure (labeled \textit{T} in Figure ~\ref{Rjet1}) began to appear in the anti-sunward direction. On September 29, a short jet (indicated by arrows in Figure~\ref{Rjet1}) in the sunward direction can be seen. Hereafter, this sunward jet feature can be detected in all our images obtained at Lulin and Calar Alto Observatory. It is interesting to note that two distinct sunward dust structures are visible after October 29 lasting until November 2. Around the same time, from November  2 to November 4, Mueller et al. (2013) also reported seeing two separate continuum features in sunward direction. Afterwards, only a single jet could be seen in the sunward direction that became fainter and fainter as Hartley 2's heliocentric and geocentric distances increased. The sunward jet features showed relatively little variation during a night but its shape and position angle slightly changed from night to night until October 11 when two distinct jet features apparently emerged from the sunward direction (Figure 5). In order to examine the existence of this extremely faint jet feature and to distinguish it from the trail of a background star, we transformed the enhance image into polar coordinates $\rho$-$\theta$ where $\rho$ is the projected cometocentric distance from the nucleus and $\theta$ is the azimuth (position angle). At several distances $\rho$ from the nucleus, we analyzed the resulting azimuthal profile. In Fig. 5 (right panel) we show the azimuthal profile at $\rho = 5,000$ km.  It can be seen that this faint jet (referred as main feature in the figure) appears on Oct. 11.76 and persists until Oct. 11.87, that is $\sim$0.7hr later (bottom panel in Fig. 5). It points towards the Sun and it does not display significant changes. On Oct. 11.84, a new faint feature appears nearly perpendicularly to the Sun-comet line. It is interesting to note that the position angle of the secondary jet is roughly the same as that of the CN jet features shown in Figure~\ref{CNjet} (pointing to the east-south direction in the top-middle panel). At first, one could think that icy grains mixed with the dust grains of this weaker jet could provide the partial fuel to the CN gas jet. However, the gas jets persist for most of a rotation period (Knight and Schleicher 2011, Samarasinha et al. 2011) and are clearly being released over an extended period of time as the nucleus rotates. Thus, the CN jets cannot mainly come from this faint jet feature if it is only active for a few hours as found here. That switching phenomenon may also be explained as a projection effect due to the comet nucleus rotation. \\

For Oct. 28 and 29, we obtained a series of images from Lulin and Calar Alto observatories that provide insight into how the sunward feature evolved throughout $\sim$1.4 rotation cycles. Representative images from these nights are shown in Fig. 6, with each panel enhanced by the Larson-Sekanina filtering method. Notice that the position angle (PA) of the Sun during these two days is near 97$^\circ$. Setting the zero phase at 11:40 UT on October 10 and using a period of 18.11 hr (see the Section 3), the rotational phase can be easily estimated in these three images (see the bottom right corner of Figure ~\ref{LotCa}). A dust jet (labeled J and marked with an arrow in Figure ~\ref{LotCa}) can be seen in the sunward direction whose shape slightly changes as the rotational phase change from 0.22 (on Oct. 28.68 UT) to 0.82 (Oct. 29.13 UT). Thirteen hours later, (rotational phase of 0.57, Oct. 29.70 UT) two dust jet features emanating in the sunward direction can be seen. One of them, labeled J1, is close to the position of Sun (PAs $\sim$85$^\circ$) and the other, labeled J2, lays at the PAs  $\rm \sim 130 ^\circ$. Owing to the similar PAs, we consider the possibility that J1 feature might have the same source region as seen  from the previous two images (Oct. 28.68 and Oct. 29.13). Under this assumption, J2 feature is new. Another possibility is J1 feature might be associated with the cometary rotational effect, i.e. local sunrise accompanied by temperature increase turns that jet on. This localized temperature difference in the regions of waist and the sun-lit end of the nucleus have been addressed by Belton et al (2013). The J2 feature which has a collimated-like shape is the persistent feature we detected on Oct. 28.68 and Oct. 29.13 although PA and shape changed between those two dates. We note that the brightness of J2 feature is higher than that of J1 feature and this higher intensity could be related to the dusty ice, or to an outburst from the surface of the comet nucleus. To understand their interrelationship better, our images need to be interpreted in the context of a larger image series that displays the time evolution of the jet structure over two or more rotational period.\\

On the tailward side, only the dust tail was readily visible starting in August, 2010. Dust tail was found to point approximately in the antisolar direction. As expected, it appears to be curved slightly counterclock-wise.
 
\subsection{The properties of dust coma}

We used Af$\rho$ (A'Hearn et al. 1984) to characterize the dust activity of the comet, the derived values acquired with broadband R-filter from April to November 2010 are presented in Figure ~\ref{afrho}. Except for the night on October 29, the average values estimated every photometric night were all measured within a projected distance of 5,000km. Notice that Af$\rho$ shows a weak dependence on the $\rho$, projected distance from nucleus, from 5,000km to 20,000km and the variation was found to be less than approx. 5-8\%. The reason why we used 5,000km for uniform radius is to reduce the influence of star trails in field of view. The $A f \rho$ values steadily increased with decreasing heliocentric distance, although there was not a noticeable increase when the second jet appeared on Oct. 11.64 UT or at the perihelion. The $A f \rho$ value on October 29.77 $\sim$ 29.85 UT increased from 155 cm to 174 cm in two hours, and at the same time the dust jet seen in Figure 6 (right panel) was more prominent on this night than on any of the other nights and a relatively weak secondary jet feature was also detected. Possible causes for this deviation might include the changes in the physical properties of the grains as they travel outward (i.e. loss of volatiles or fragmentation), the action of solar radiation pressure modifying the straight trajectories of small particles inside the field of view, or a long-lasting population of large particles (Schleicher et al. 1998). Furthermore, the power law index of the $r_h$ dependence for the dust,  Af$\rho$  (5,000 km), is $-3.75 \pm 0.45$ before perihelion and is $-3.44 \pm 1.20$  post-perihelion. This is result is completely consistent with Knight and Schleicher's (2013) when using A($\theta$)f$\rho$.  \\

The derived  Af$\rho$ values for the narrowband filter can be taken to estimate the color of the cometary dust \citep{jew87} as the normalized gradient of the  Af$\rho$ product between the blue (BC,$\lambda_{0}$= 4,430 \AA) and red (RC,$\lambda_{0}$= 6,840 \AA) continuum filters. The dust color can be converted to a percentage of reddening per 1,000 {\AA} and is defined by the following relation: 

\begin{equation}\label{Afrho_formula}
color = \frac{RC_{Af\rho}-BC_{Af\rho}}{6840-4430}\frac{2000}{RC_{Af\rho}+BC_{Af\rho}}
\end{equation}

\begin{figure}[p!]
\begin{center}
\includegraphics[width=0.7\textwidth]{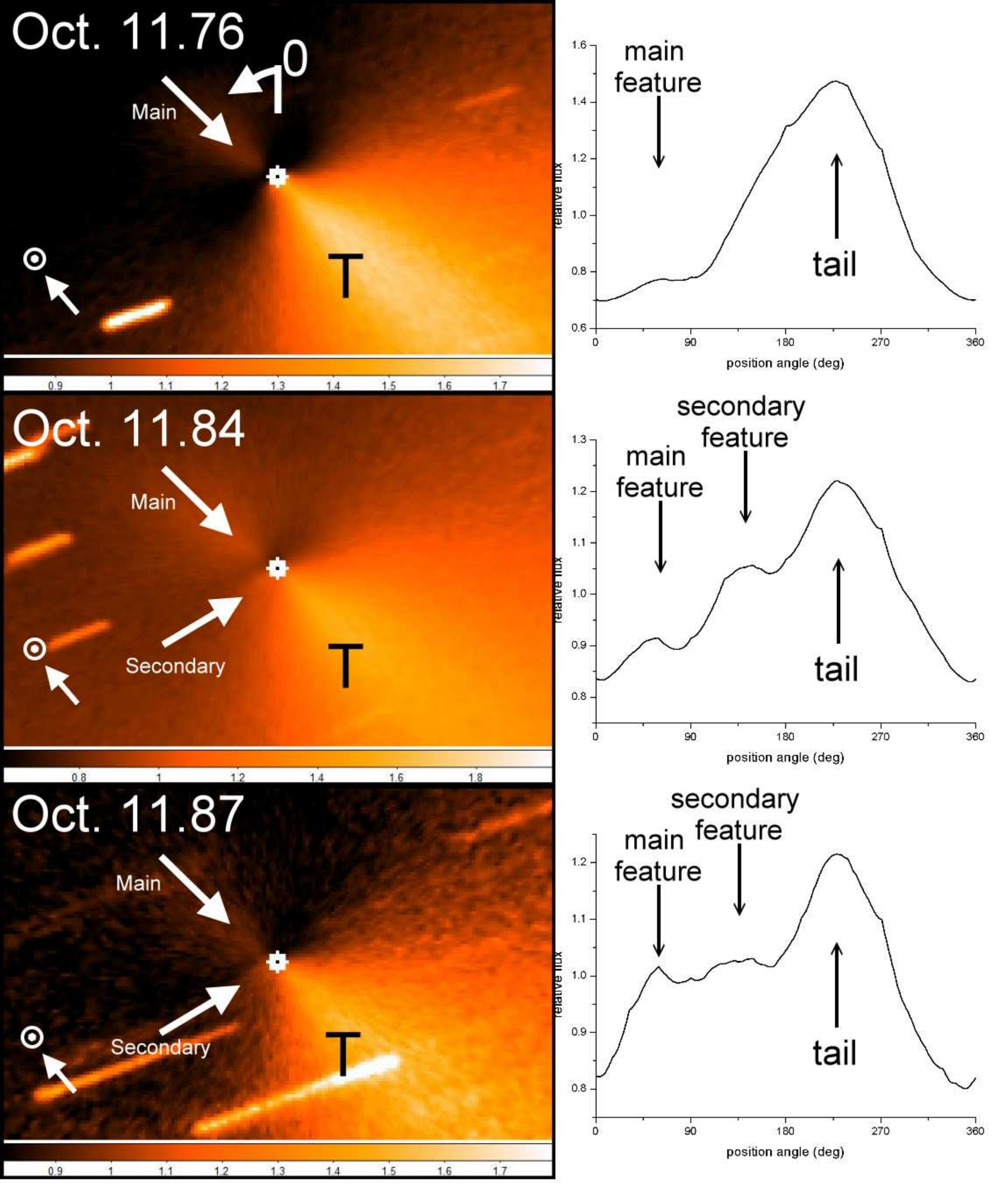}
\caption{
	\label{Rjet2}
	Dust jet features and tail enhanced by an azimuthal median profile (left panels) and the corresponding azimuthal profiles obtained at $\rho$ $\sim$ 5,000 km (right panels). A sunward dust jet feature is revealed in broadband the R-filter on Oct. 11.76UT (top panel). Two faint dust jet features are detected using both broadband R-filter (middle panel) and narrowband red-continuum filter (bottom panel) on Oct. 11.84 UT and 11.87 UT, respectively. The straight jet pointing towards the Sun (main feature) and the weaker one pointing nearly perpendicular to the Sun-nucleus direction (secondary feature) are marked in the graphs. Position angle is measured from north (up) in the counterclockwise direction (top-left panel).  In the left panels, all images are centered on the nucleus, arrows with the indicating the jets and T for the dust tail, North is up, East is to the left, the field of view is 2.92' $\times$ 1.94', corresponding to 9,200 $\times$ 6100 km at the comet distance, the Sun symbol and the corresponding  arrow indicate the projected direction towards the Sun.}
\end{center}
\end{figure}

\begin{figure}[p!]
\begin{center}
\includegraphics[width=\textwidth]{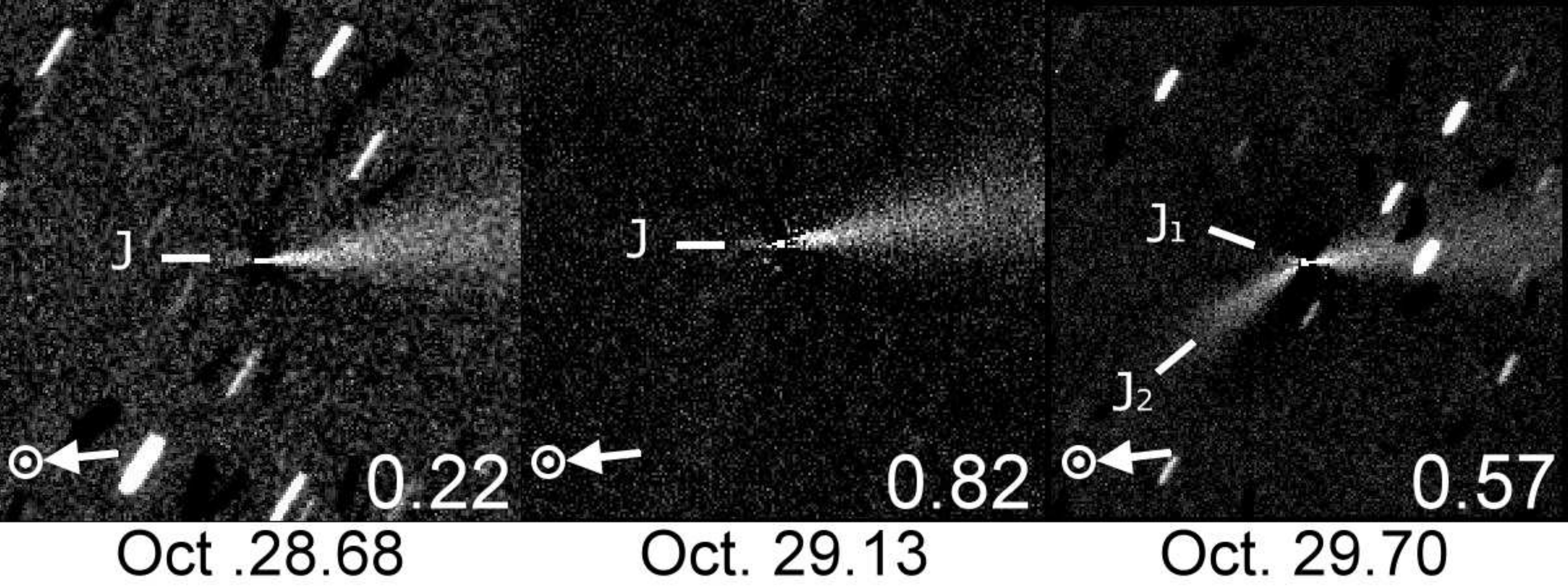}
\caption[]{
	\label{LotCa}
	Time sequence of images of the comet 103P/Hartley 2 acquired from Lulin (left and right) and from CA (middle) observatories. Dust jet features are enhanced by Larson-Sekanina filtered. The rotational phase is given at the bottom right corner of each image.  North is up, East is to the left. The field of view is 3.8' $\times$ 3.8' and all images are centered on the nucleus.  J, J$_1$ and J$_2$ refer to the jets and the Sun symbol and arrow indicate the projected direction towards the Sun.
	}

\end{center}
\end{figure}

\begin{figure}[p!]
\begin{center}
\includegraphics[width=\textwidth]{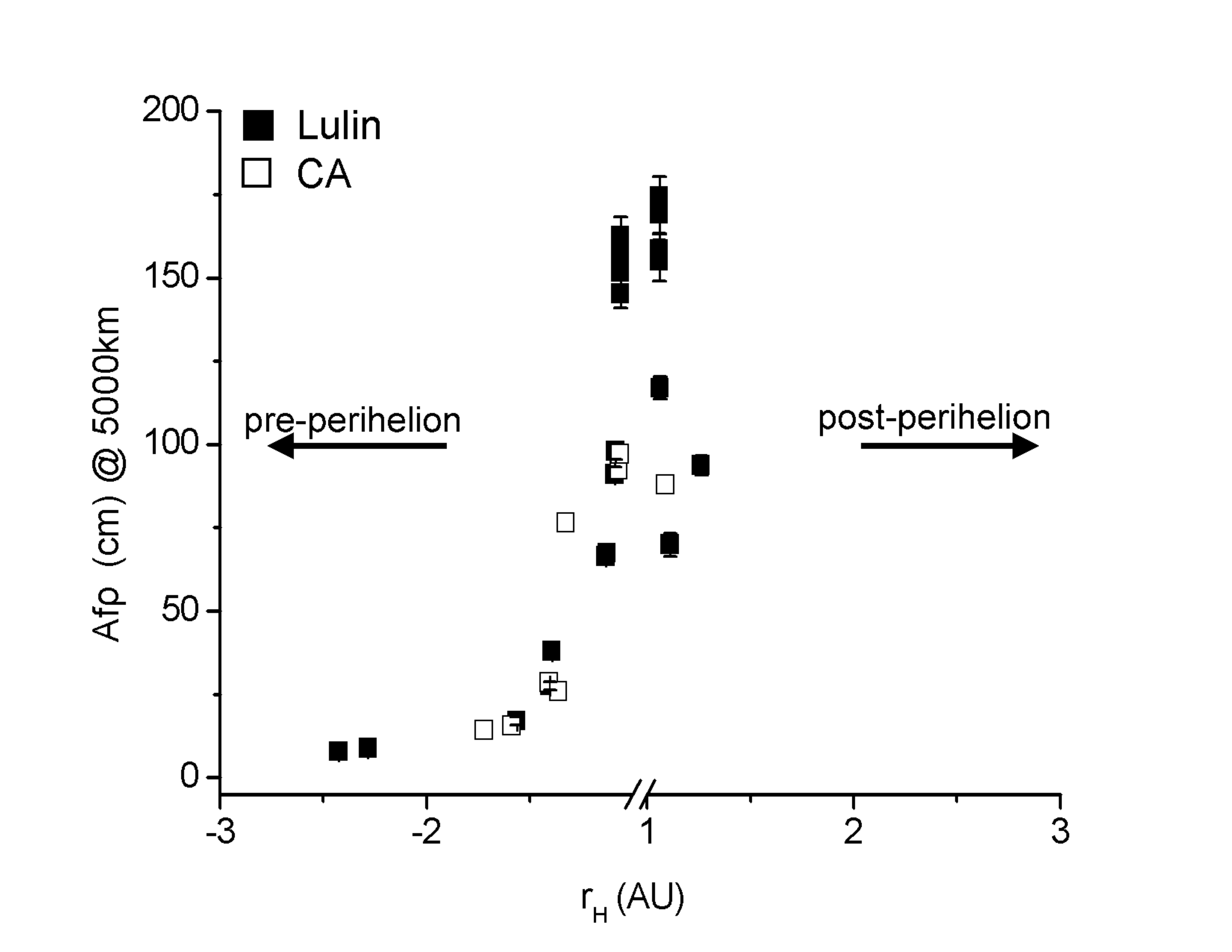}
\caption{
	\label{afrho}
	Af$\rho$ variation as a function of heliocentric distance both pre- and post-perihelion. Filled squares indicate the results obtained from Lulin observatory (LOT) and opened squares pertain to the data from Calar Alto observatory. "//" is referred to the break heliocentric distance from $\rm -0.1~AU$ (pre-perihelion) to 0.99 AU (post-perihelion). The error bars are not clearly seen in this figure because they are lower than 5\%.
	}

\end{center}
\end{figure}

\begin{figure}
%%\epsscale{.80}
%%\plottwo{f8a.eps}{f8b.eps}
\plottwo{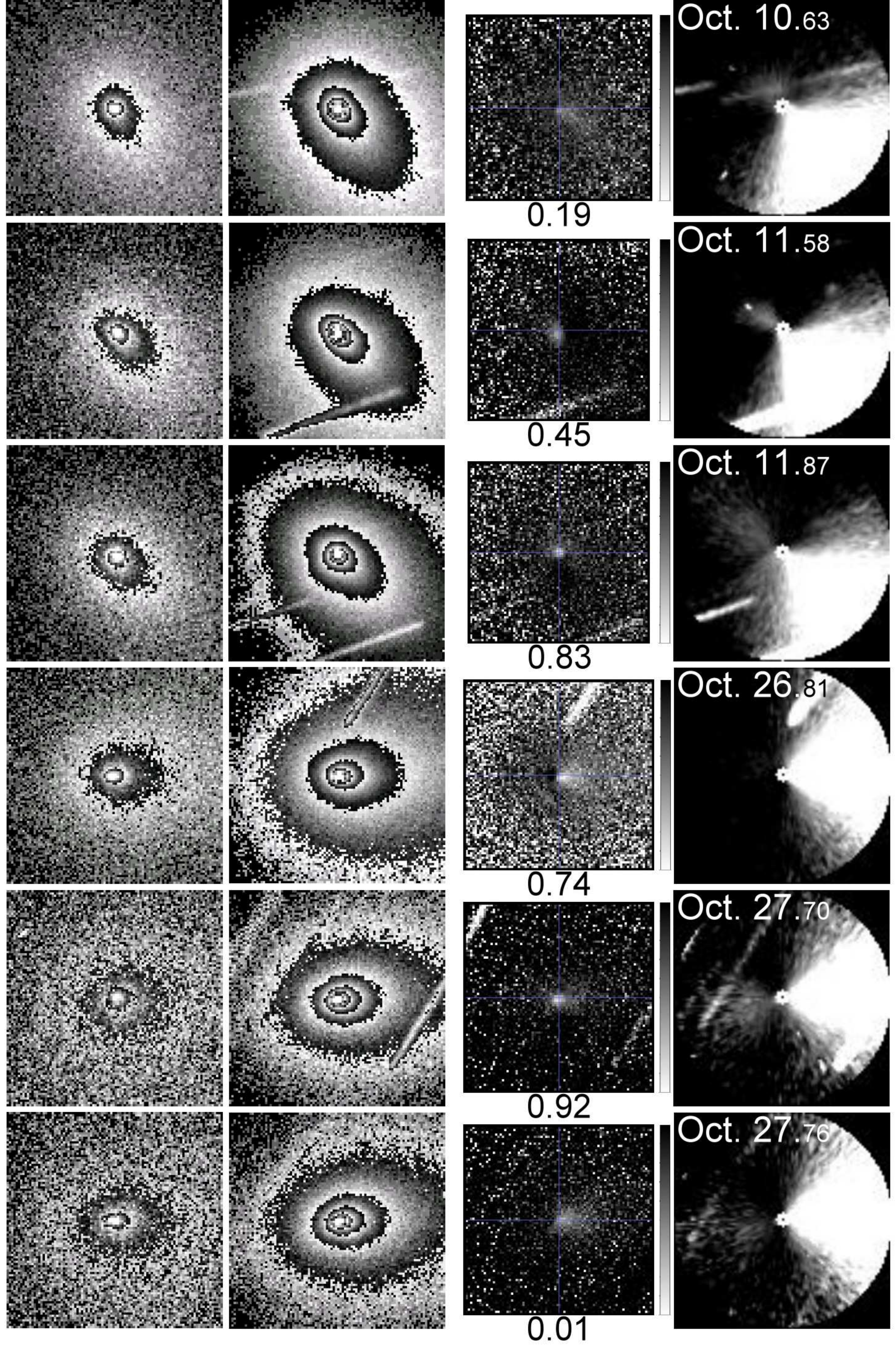}{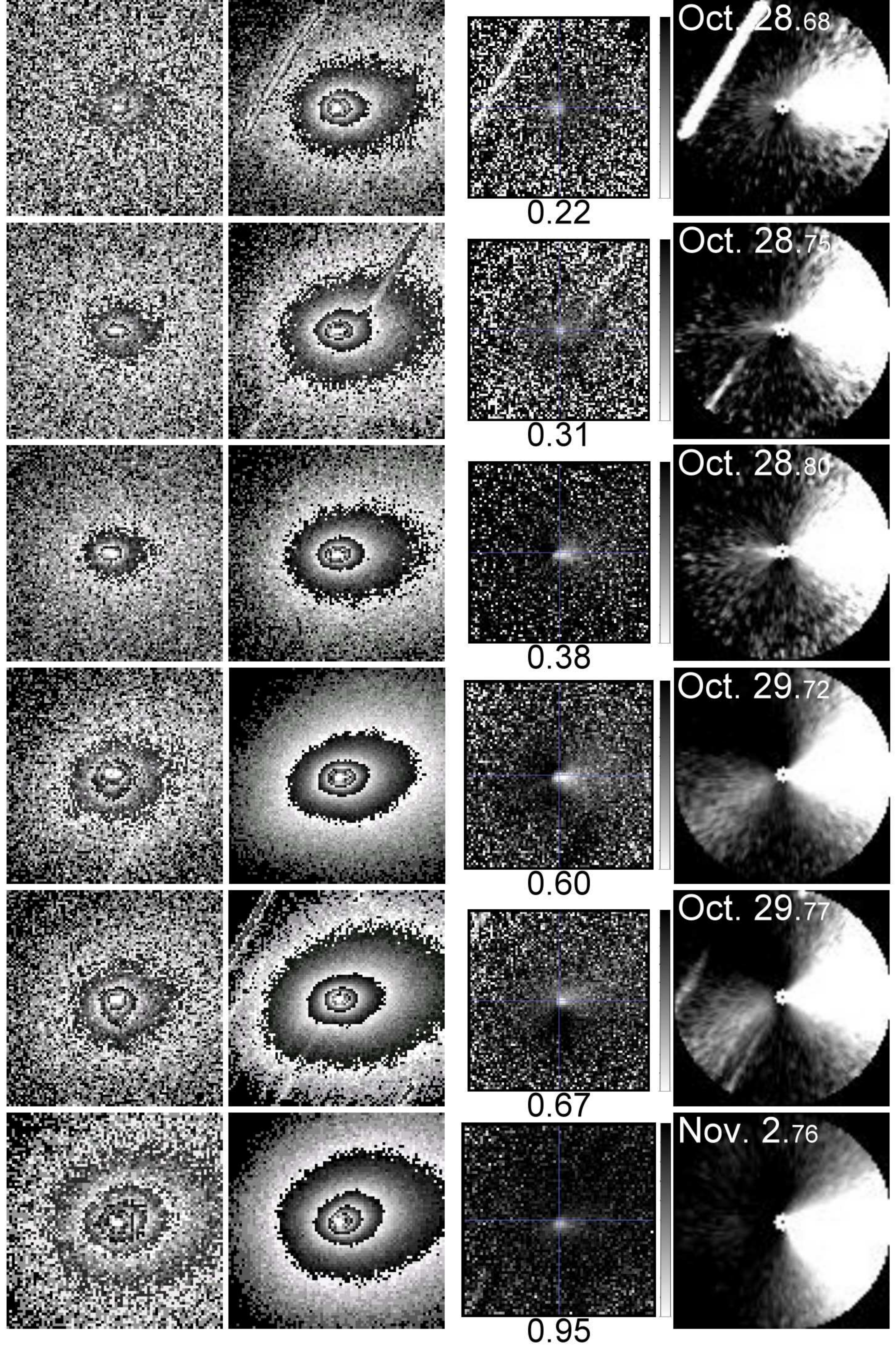}
\caption{Jet activity and dust color of the coma of 103P/Hartley 2. The first two columns are the images acquired with blue continuum filter centered at 443 nm (BC) and with red continuum filter centered at 684nm (RC), respectively. The third column shows the dust reddening computed with equation 1. The color bar stretches from 0 (black) to 50$\%$ (white) / 100 nm. The fourth column displays the ring-masking images obtained by subtracting the RC images from an image generated with the azimuthal average profile. North is up, East is to the left. The field of view is about 40" $\times$ 40", corresponding to 1,800 km$\sim$2,200 km at the comet distance depending on the different comet heliocentric distance. \label{fcolor}}
\end{figure}

The summarized results in Table~\ref{tcolor} indicate that the averaging dust color within the  innermost 5,000 km of the coma did not appear to vary significantly with heliocentric distance. This behavior of the averaging dust color seems to indicate that the innermost coma do not introduce significant changes on the size distribution and/or overall properties of dust grains. As we found a jet feature that switches on and off from our images in Figure~\ref{Rjet1} to Figure~\ref{LotCa}, we analyzed the entire flux-calibrated images acquired with BC and RC narrowband filters instead of integrating whole flux in the innermost 5,000 km. The resulting two-dimensional dust color map can be seen in Figure~\ref{fcolor} (the third column). Figure~\ref{fcolor} displays the dust coma of comet 103P/Hartley 2 from October 10 to November 2 imaged in BC and RC narrowband filters (first two column), the dust color map (the third column) and azimuthal median profile subtracted RC filter images (the fourth column) that displays the jet activity in the dust coma. The data here presented pertaining to October and November give an extremely reddened dust, with a normalized color $\sim$ 30-45 \%, within a radius of $\sim 50-100$~km measured from the optocenter of the images. This red dust could be associated with strong jet activity. The sunward jet feature might give rise to higher dust abundances at closer cometocentric distances (i.e. near the optocenter). These dust grains are initially large with a reddening of $\sim 30-40$\%/1,000\AA, while travelling out they split up and show bluer at $\sim 500$ km with a dust reddening of $\sim$ 10-15 \%. In comparison with tailward direction, the color variation is 5\% to 10\%. The decrease in the dust reddening means that the optical properties of the dust grains change as the dust grains move outward or this blueing of the dust could be also associated with an outburst (Bonev et al., 2002). A possible explanation for color variation is that the larger dust grains mixed with the icy grains dominate the scattering behavior at close distance around the nucleus. When these larger dust grains move outwards, they break up or sublimate into the small sub-micron particles resulting in a bluer continuum due to their smaller sizes (Lara et al. 2011b). 

\section{Summary}
We observed the comet Hartley 2 at the Lulin Observatory in Taiwan, the Calar Alto Observatory in Spain, and the Beijing Astronomical Observatory in China,  from April to December, 2010 using both broadband and narrowband filters, and long-slit spectrophotometry. The results are summarize below.\\
1. CN morphology: The processed CN images revealed two asymmetric jet features in the coma of comet Hartley 2. The CN jet features detected in the images here presented did not show the sprial-like structure seen by other authors in earlier date due to different observing geometry. One of these CN jet features always shows a higher intensity than the other, possibly because it is facing towards the Earth.\\
2. Gas production rates: Our Lulin, BAO and CA results show that there is no significant variation of Q(CN) from mid-October to early-November. The power law slopes of the heliocentric distance of the gas production rate of CN and C$_{2}$ are $-4.57$ and $-4.84$ before perihelion and $-3.21$ and $-3.42$ after perihelion. The average C$_2$-to-CN production rate ratio is 0.7$\sim$ 1.5 which places 103P/Hartley 2 as a "typical" in terms of $\rm C_2$ enrichment.\\
3. Dust morphology: The sunward jet feature was first detected in images acquired at the end of September, 2010. This sunward jet seems to be non-permanent. Instead, morphology varies with time and two distinct jet features are found on October 11 and after October 29 until November 2. \\
4. Af$\rho$ and dust color: The power law $r_h$ dependence of the dust production rate,  Af$\rho$  (5,000 km), is $-3.75\pm0.45$ before perihelion and $-3.44\pm1.20$ during post-perihelion. The higher dust reddening is found around the optocenter and could be associated with a stronger jet activity. The dust color is getting bluer outwards along the sunward jet which implies that the optical properties of the dust grains change with $\rho$. The average dust color did not appear to vary significantly when the heliocentric distance decreased to perihelion.

\acknowledgments
This work was based on observations obtained at Taiwan's Lulin Observatory. We thank the staff members and Yu-Chi Cheng for their assistances with the observations. We greatfully acknowledge valuable discussions with the referee. The research was supported by project AyA2009-08011 of the Ministerio de Ciencia e Innovacion. Zhong Yi Lin acknowledges a post-doctoral grant awarded by the Junta de Andalucia through project number P07-TIC-274. This work was also supported by grant number NSC 99-2923-M-008-002-MY3 for the Formosa Program (NSC-CSIC).
%%This work has made use of NASA's Astrophysics Data System.  It 
%%also benefitted tremendously from \citet{latexguide}.

% Bibliographic references with the natbib package:
% Parenthetical: \citep{Bai92} produces (Bailyn 1992).
% Textual: \citet{Bai95} produces Bailyn et al. (1995).
% An affix and part of a reference:
%   \citep[e.g.][Ch. 2]{Bar76}
%   produces (e.g. Barnes et al. 1976, Ch. 2).-

%%\bibitem{Jones et al.(1990)}

%% Use the plainnat style for ``Icarus'' mode to display DOI numbers
%% among other things.  However, revert to the Elsevier elsart-harv
%% mode for ``Elsevier'' mode.
%%\bibliographystyle{plain}

%% --Tables-- 

\begin{deluxetable}{cccccccccc}
%%\tabletypesize{\scriptsize}
\rotate
\tablewidth{0pt}
\tablecaption{Log of observations\label{ObsLog}}
\tablehead{
\colhead{Date}           & \colhead{UT}      &
\colhead{$\rm r_h$}      & \colhead{$\Delta$}  &
\colhead{P.A.}          & \colhead{$\alpha$}    &
\colhead{Pix scale}  & \colhead{Data}  &
\colhead{Obs.} & \colhead{Sky} } 
\startdata 
April 24	&19:56-20:17 &2.424 & 2.363 &256.2 & 24.2 & 884.4 & R & Lulin& Phot.\\
May 11 	&18:39-18:48 &2.283 &2.026 &252.1 &26.3 &758.2 & R & Lulin	& Phot.\\
May 15 	&18:58-19:10 &2.249 &1.948 &251.0 &26.7 &729.1 & R & Lulin	& Part. cloudy \\
May 16 	&19:07-19:36 &2.240 &1.928 &250.8 &26.8 &721.6 & R & Lulin	& Part. cloudy \\
May 20 	&18:34-18:56 &2.206 &1.851 &249.7 &27.1 &692.7 & R & Lulin	& Part. cloudy \\
July 14 	&17:01-18:40 & 1.721 & 0.924 &226.9 &29.0 &345.8 & R & Lulin	& Part. cloudy \\
July 14 	&23:00-23:49 & 1.719 & 0.921 &226.7 &29.0 & 354.0 & R & CA	& Phot.\\
July 22 	&01:25-02:36 & 1.656 & 0.825 &221.7 &29.0 & 316.1 & R & CA	& Part. phot.\\
July 30 	&02:22-02:55 & 1.585 &0.725 &215.2 &29.2 &278.7 & R & CA	& Phot.\\
August 1 	&17:30-18:42 &1.562 &0.694 &212.9 &29.3 &259.7 & R & Lulin	& Phot.\\
August 19 	&14:00-20:15 &1.409 &0.503 &195.5 &31.1 &188.3 & R & Lulin	& Part. cloudy \\
August 20 	&01:33-01:44 & 1.406 &0.500 &195.1 &31.1 &192.2 & R & CA	& Phot.\\
August 20 	&13:17-20:01 &1.402 &0.494 &194.5 &31.2 &184.9 & R & Lulin	& Phot.\\
August 21 	&19:27-20:43 &1.392 &0.483 &193.3 &31.4 &180.8 & R & Lulin	& Phot.\\
August 25 	&22:46-00:23 &1.359 & 0.445 &189.3 &32.3 &174.5 & R,S & CA	& Phot.\\
August 29	&18:37-19:09 &1.329 &0.412 &185.6 &33.3 &154.2 & R & Lulin	& Part. cloudy \\
September 2 	&02:23-02:27 & 1.304 & 0.384 &182.7 &34.2 &147.6 & R,S & CA	& Part. phot.\\
September 14 	&00:35-00:57 & 1.219 & 0.293 &175.1 &38.3 &109.2 & R,S & CA	& Phot.\\
September 29 	&14:45-20:51 &1.130 &0.192 &182.1 &44.2 &71.9 &R& Lulin	& Phot.\\
September 30	&13:55-20:06 &1.125 &0.186 &183.8 &44.6 &69.6 &R& Lulin	& Phot.\\
October 2 	&17:55-18;05 &1.116 &0.175 &188.3 &45.3 &65.5 &R& Lulin	& Part. phot.\\
October 3 	&15:21-15:30 &1.112 &0.170 &191.1 &45.6 &63.6 &R& Lulin	& Part. phot.\\
October 9        &14:40-17:55 &1.090 &0.143 &211.7 &47.7 & &S & BAO	& Part. phot.\\
October 10 	&12:32-18:45 &1.087 &0.140 &215.9 &48.0 &52.4 &R+N & Lulin	& Phot.\\
October 11 	&15:08-19:00 &1.083 &0.136 &220.3 &48.4 & &S & BAO	& Phot.\\
October 11 	&13:05-21:08 &1.083 &0.136 &220.3 &48.4 &50.9 &R+N & Lulin	& Phot.\\
October 15 	&00:28-02:18 & 1.077 &0.129 &230.9 &49.4 &48.8 & R,S & CA	& Part. cloudy \\
October 18 	&04:38-05:10 &1.068 & 0.122 &248.1 &51.4 &46.9 & R,S & CA	& Phot. \\
October 19 	&22:32-23:15 & 1.065& 0.121 &254.4 &52.5 &46.5 & R,S & CA	& Part. phot.\\
October 22 	&02:39-03:13 &1.062 &0.121 &261.0 &53.8 &46.5 & R,S & CA	& Phot.\\
October 25 	&03:53-04:38 & 1.060 &0.125 &268.6 &55.6 &47.6 & R & CA	& Part. cloudy \\
October 25 	&18:09-18:39 &1.059 &0.126 &269.8 &55.9 &47.2 &R+N & Lulin	& Phot.\\
October 26 	&16:09-21:24 &1.059 &0.128 &271.8 &56.4 &47.9 &N & Lulin	& Phot.\\
October 27 	&16:24-26:56 &1.059 &0.130 &273.7 &56.5 &48.7 &N& Lulin	& Phot.\\
October 28 	&17:50-21:28 &1.059 &0.132 &275.3 &57.4 &49.4 &N & Lulin	& Phot.\\
October 29 	&03:06-03:20 & 1.059& 0.133 &275.9 &57.5 &51.1 & R & CA	& Part. cloudy\\
October 29 	&16:24-21:00 &1.059 &0.135 &276.9 &57.7 &50.5 &R+N & Lulin	& Phot.\\
November 1 	&19:14-21:12 &1.060 &0.145 &280.9 &58.6 &54.3 &R & Lulin	& Part. cloudy \\
November 2 	&17:58-20:31 &1.061 &0.150 &282.0 &58.7 &56.1 &R+N & Lulin	& Phot.\\
November 3 	&16:04-16:28 &1.061 &0.149 &281.9 &58.7 &55.7 &R & Lulin	& Part. cloudy \\
November 5 	&18:07-21:09 &1.064 &0.157 &284.2 &58.8 &58.8 &R & Lulin	& Part. cloudy \\
November 5 	&01:25-05:45 & 1.065 &0.162 &285.4 &58.8 & 62.3 & R,S & CA	& Part. phot.\\
November 16 	&01:18-01:47 & 1.090 &0.210 &294.0 &56.1 &80.3 & R,S & CA	& Part. phot.\\
November 21 	&19:07-19:37 &1.112 &0.239 &298.8 &53.2 &60.3 &R & Lulin	& Phot. \\
December 2 	&16:24-18:48 &1.116 &0.295 &308.9 &46.4 &74.5 &R & Lulin	& Part. cloudy \\
December 17 	&17:48-20:58 &1.262&0.377 &326.2 &36.4 &95.2 &R & Lulin	& Phot. \\
December 26	&23:08-00:12 & 1.323& 0.427 &338.0 &31.3 &163.7 & R,S & CA	& Phot. \\
\enddata
\tablecomments{$\Delta$ and $r_h$ are the geocentric and heliocentric distances in AU; P.A. is the extended Sun-target radius vector as seen in the observer's plane-of-sky, measured from North towards East. $\alpha$ is the phase angle (Sun-comet-observer angle). R is the broad band filter, N is the cometary filter set and S refers to long-slit spectra.}
\end{deluxetable}	

\clearpage
\begin{deluxetable}{ccccccccc}
\tabletypesize{\footnotesize}
\rotate
\tablewidth{0pt}
\tablecaption{Gas production rates of comet 103P/Hartley 2. \label{Qs}}
\tablehead{
\colhead{Date} & \colhead{UT}  & \colhead{Observatory}      & \colhead{Aperture}  &
\colhead{CN}          & \colhead{C$_{2}$}    & \colhead{C$_{3}$}  & \colhead{NH$_{2}$}  &
\colhead{Q(C$_{2}$)/Q(CN)} } 
\startdata 
Sept. 14&00:35-00:57&CA	&& 5.82$\pm$0.13	& 7.28$\pm$1.03	& 0.37$\pm$0.05 	& X &  \\ 
Oct. 9&14:40-17:55&BAO	&&21.9$\pm$4.78	&23.4$\pm$5.24	&4.39$\pm$0.94	&44.4$\pm$9.70&1.07$\pm$0.11\\
Oct. 10&12:32-18:45 &Lulin	&4.01&6.57$\pm$0.93	&	X		&	X		&	X	&	\\
Oct. 11&15:08-19:00&BAO	&&21.2$\pm$3.57	&24.3$\pm$4.10	&4.37$\pm$0.74	&44.0$\pm$7.38&1.15$\pm$0.11\\
Oct. 11&13:05-21:08&Lulin	&4.00&9.32$\pm$1.37	&	X		&	X		&	X	&	\\
Oct. 18&04:38-05:10 &CA	&&  21.4$\pm$3.56	& 14.8$\pm$1.57 	& 1.34$\pm$0.02 	& 35.2$\pm$1.41  & 0.69$\pm$0.04\\
Oct. 18&04:38-05:10 &CA	&&  21.2$\pm$1.39	& 15.1$\pm$3.99 	& 1.33$\pm$0.002	& 46.9$\pm$0.97  & 0.71$\pm$0.29 \\
Oct. 22&02:39-03:13 &CA	&&  20.2$\pm$2.90	& 24.6$\pm$1.57 	& 1.56$\pm$0.05	& X &1.22$\pm$0.05\\
Oct. 25&18:09-18:39&Lulin	&3.62&18.2$\pm$3.1 	&	X		&	X		&	X	&\\
Oct. 26&16:09-21:24&Lulin	&3.66&20.1$\pm$5.4	&	X		&	X		&	X	&\\
Oct. 27&16:24-26:56&Lulin	&3.82&12.7$\pm$2.7 	&	X		&	X		&	X	&\\
Oct. 28&17:50-21:28&Lulin	&3.98&18.9$\pm$3.2	& 27.8$\pm$5.3	&	X		&	X	&1.47$\pm$0.17\\
Oct. 29&16:24-21:00 &Lulin	&3.99&21.8$\pm$3.6 	& 31.6$\pm$5.4	&	X		&	X	&1.45$\pm$0.15\\
Nov. 2&17:58-20:31 &Lulin	&4.17&20.3$\pm$3.1 	& 22.0$\pm$3.2	&	X		&	X	&1.07$\pm$0.10\\
Dec. 26&23:08-00:12 &CA	&&7.54$\pm$0.13 & 9.64$\pm$1.6 & 0.29$\pm$0.05 &  X   &1.29$\pm$1.23\\
\enddata
\tablecomments{The unit of aperture size in log-scale is km. The unit in all gas production rates is $10^{24}$ molecules/s.}
\end{deluxetable}	

\clearpage

\begin{table}
\begin{center}
\caption{The dust color averaged within the innermost 5,000 km of the coma \label{tcolor}}
\begin{tabular}{ccc}
\tableline\tableline
Date & color (\% / 100 nm) \\
\tableline
October 10	&9.40$\pm$0.82\\
October 11	&11.50$\pm$0.934\\
October 25	&5.06$\pm$0.78\\
October 26	&6.36$\pm$0.82\\
October 27	&6.43$\pm$0.87\\
October 28	&11.07$\pm$0.83\\
October 29	&10.01$\pm$0.83\\
November 2	&5.56$\pm$0.88\\
\tableline
\end{tabular}
\end{center}
\end{table}

\clearpage

 %%--Figures-- %%

\clearpage

\clearpage

\clearpage

\clearpage

\clearpage

\clearpage

\end{document}